\documentclass[letterpaper,twocolumn,10pt]{article} 

\usepackage{usenix-2020-09} 
\usepackage{graphicx} 
\usepackage{booktabs} 
\usepackage{multirow} 
\usepackage{amsmath} 
\usepackage{xcolor} 
\usepackage{enumitem}

\newif\ifshowcomment
\showcommenttrue

\ifshowcomment
\newcommand{\xiaohu}[1]{\textsf{\color{black}{#1}}}
\newcommand{\todo}[1]{\textcolor{red}{[TBD: #1]}}
\else
\newcommand{\todo}[1]{}
\newcommand{\xiaohu}[1]{}
\fi

\newcommand{\ddparagraph}[1]{\vspace {1pt}\noindent\textbf{#1}}

\begin{document}


\title{\bf \Large HBF-Sim: An Extensible HBF Simulator for Large-scale GPU Memory Systems}


\author{
{\rm Yaqi Li}$^{1,2}$, 
{\rm Jing Wang}$^{1}$, 
{\rm Junfeng Wang}$^{1}$, 
{\rm Long Yang}$^{1,2}$, 
{\rm Han Yan}$^{1}$, 
{\rm Xiaohu Chai}$^{1}$, 
{\rm Liang Shi}$^{*1,2}$ \\[1ex]
$^{1}$\textit{East China Normal University}  \qquad
$^{2}$\textit{Shanghai Innovation Institute}
}
\date{}
\maketitle

\begin{abstract}
High-bandwidth flash (HBF) is introduced to address the memory wall, which can co-package a dense NAND stack with the GPU, targeting the performance gap between near-accelerator bandwidth and flash density. HBF, however, is neither a large HBM nor a fast NVMe SSD. Its usable bandwidth depends on how GPU cache-line requests map onto NAND pages, how concurrency spreads across channel-affine die sets, and how media management interacts with the GPU memory pipeline. To our knowledge, existing GPU, SSD, or HBF simulators cannot faithfully model this behavior.

We present HBF-Sim, an extensible, reusable, and faithful HBF simulator integrated with simulated GPUs. It closes the loop between GPU issue limits, device queuing, and NAND behavior in one end-to-end request path. HBF-Sim separates a GPU–HBF interaction controller from page-based parallel stack storage, and it models the full GPU–HBF request path. It provides an MSHR-based address mapping table that merges cache-line requests into page-based operations, as well as a page-based multi-stack flash manager for highly parallel reads and writes. Validation tests and device-level microbenchmarks expose performance bottlenecks caused by limited channel distribution and resource conflicts, offering concrete guidance for next-generation HBF architectures. 

\end{abstract}

\section{Introduction}

Memory-hungry workloads such as LLM training and agentic inference are increasingly bounded by the limited capacity of on-chip HBM and DDR memory~\cite{ai_memory_wall}. Model weights are largely read-only and fixed in size~\cite{vllm}, whereas the key–value (KV) cache grows with context length and batch size. During decoding, insufficient HBM capacity forces systems either to evict reusable KV entries or to shard models across more GPUs. Recent work has explored CPU DRAM and flash as extensions to GPU memory~\cite{flexgen,llminflash}, but these remote memory tiers remain constrained by PCIe bandwidth, access latency, and software-stack overhead.

High-bandwidth flash (HBF) offers a promising way to alleviate the GPU memory wall. As a package-local memory tier for GPUs, it provides 512 GiB of density and multi-TB/s interface bandwidth at a fraction of HBM's cost. A recent Open Compute Project (OCP) specification~\cite{ocp_hbf_v07}, proposed by SK hynix and SanDisk, integrates a dense NAND stack co-packaged with the GPU and served by up to 16 independent UCIe host channels, yielding an aggregate interface bandwidth of 3.072\,TB/s. HBF is thus positioned as a "memory-interface flash" tier, offering near-accelerator latency and bandwidth together with flash density and cost.

However, simulating HBF behavior is not easy. An HBF-enhanced architecture is neither a large HBM nor a fast NVMe SSD. The main difficulty is to leverage the available bandwidth, which depends on several factors. In detail, it may depend on how 64 B GPU cache-line requests map onto 4 KiB NAND pages, how request concurrency spreads across channel-affine die sets, and how 4 KiB write aggregation and backpressure interact with the GPU's memory pipeline. It relates to media management, which prohibits the garbage-collection machinery that every SSD simulator assumes.
Unforturenately, existing platform ignore these interactions. GPU simulators (GPGPU-Sim~\cite{gpgpusim}, Accel-Sim~\cite{accelsim}) abstract off-chip storage as DRAM. SSD simulators and emulators (MQSim~\cite{mqsim}, FEMU~\cite{femu}, Cylon~\cite{cylon}) implement NVMe/CXL block-device semantics. Recent HBF studies (e.g., H${}^3$~\cite{h3}, HAVEN~\cite{haven}, TileLens~\cite{tilelens}, FlashAccel~\cite{flashaccel}, DASH~\cite{kim2026capacityscalablemoellm}, and Li et al.~\cite{hbfsucks}) build on internal or application-specific models. Their publicly available descriptions do not establish support for the full set of GPU–HBF request and completion semantics studied in this work.


This paper presents HBF-Sim, an extensible, reusable, and more faithful HBF simulator integrated with Accel-Sim/GPGPU-Sim. It models the full GPU–HBF request path and is designed to study hardware and system design. HBF-Sim explicitly separates the GPU–HBF interaction controller, which sits above the base die and owns granularity alignment, address mapping, and memory/flash management, from the page-based parallel stack storage, which owns coalescing-based data storage and data retrieval. The GPU drives HBF through ordinary load/store interfaces, while a detailed per-request trace exposes both request state transitions and accounting. We make the following contributions:

\begin{itemize}

\item \textbf{An Open-source and Scalable Simulation Framework.} 
Our HBF-Sim mainly consists of two parts,  GPU-HBF interaction controller for data alignment and page-based flash manager for multi-stack parallelism. 
We implement HBF-Sim as an extension to Accel-Sim with a configurable GPU-HBF request/ completion path and release it as open source (\S\ref{sec:design}).
\item \textbf{High-Performance Flash Access Strategies.} We propose an MSHR-based address mapping table that efficiently merges GPU cache-line requests into page-based HBF operations (\S\ref{sec:interaction}). We further design a page-based, multi-stack flash management method that supports highly parallel reads and writes (\S\ref{sec:flashmanager}).

\item \textbf{Flexible System Extension and Performance Optimization.} Our simulator is built on an extensible architecture that supports flexible configuration of storage capacity, bandwidth, NAND timing, placement, mapping, and controller limits (\S\ref{sec:design}). Users can readily extend this architecture and plug in their own optimizations (\S\ref{sec:cases}).

\item \textbf{Insights for Future HBF-Enhanced System Design.} Through validation tests and device-level microbenchmarks, we measure fine-grained workload behavior and cost (\S\ref{sec:evaluation}). Our observations identify performance bottlenecks caused by limited channel distribution and resource conflicts, which inspire potential designs for next-generation HBF architectures.


\end{itemize}

\section{Background}

\subsection{HBF Device Semantics}

\begin{figure}[h]
\setlength{\abovecaptionskip}{-15pt}
  \setlength{\belowcaptionskip}{-5pt}
\centering
\includegraphics[width=\columnwidth]{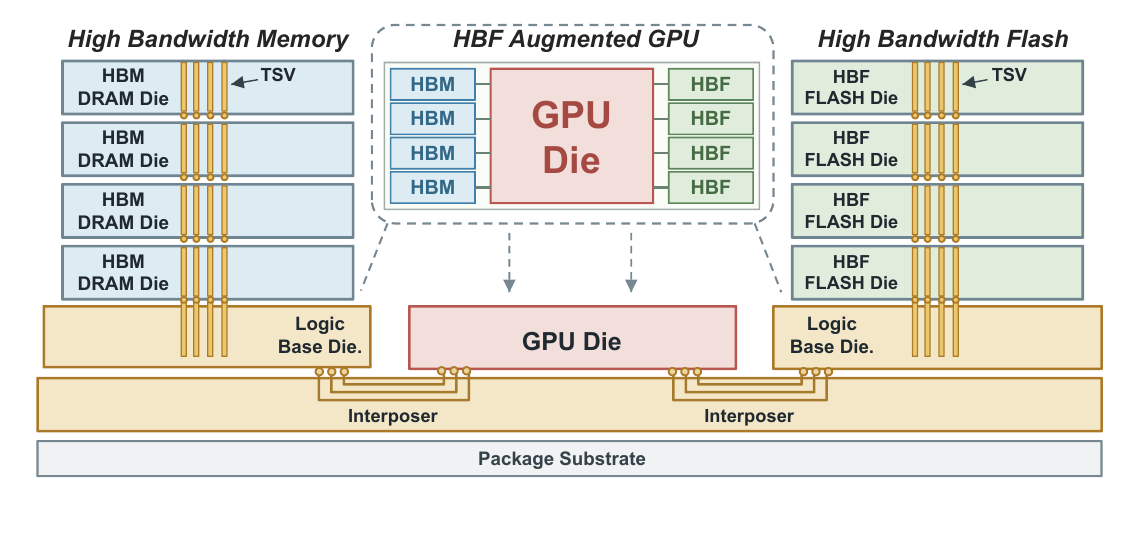}
\caption{\textbf{Overview of an HBF-augmented GPU package.}
The inset shows HBM and HBF stacks beside the GPU; cross-section shows
DRAM and NAND dies, TSVs, microbumps, logic/base dies, the interposer,
and package substrate.}
\label{fig:hardware}
\end{figure}

HBF is defined as a new storage tier that bridges the gap between HBM (High Bandwidth Memory) and SSDs (Solid-State Drives), aimed at addressing the “memory wall” challenge posed by the era of large AI models.
SK Hynix and SanDisk have jointly released the first High Bandwidth Flash (HBF)standard specification  (version 0.7.0)~\cite{ocp_hbf_v07} and made it available through the OCP (Open Compute Project). 
HBF is exposed to the host through a memory-like interface rather than as a block device.
The cube provides up to 16 independent host channels, each bound to a fixed set of NAND dies and a local address space.
Because requests on one channel cannot access resources owned by another, placement directly determines how much parallelism the device can deliver.
The aggregate interface bandwidth reaches 3.072\,TB/s across all channels, but that bandwidth is only attainable when requests are mapped to the appropriate channel-local resources.

HBF differs from conventional memory in its media semantics.
Host reads are 64B granular, while the NAND HBFs operates at 4\,KiB page granularity.
A host read must therefore be aligned to a single NAND page, which may be served from a page buffer when ordering constraints permit.
Writes are non-posted: the controller accumulates a full 4\,KiB page before programming NAND, and completion occurs only after programming finishes.
Pages within a block are written sequentially from page 0.

Unlike an SSD, there's no device-managed garbage collection on HBF. Instead, wear management is handled on the host side through zone remapping, and no background valid-page migration is performed. Thus, SSD-style FTLs serve only as a reference, not as the primary model for HBF behavior.

Table~\ref{tab:tiers} compares HBF with HBM, CXL-flash, and NVMe SSDs.
Compared with HBM, HBF offers much larger capacity but retains flash-like latency and page granularity.
Compared with CXL-attached flash and NVMe SSDs, HBF keeps a memory-like load/store interface while moving the flash device onto the GPU package, thereby reducing host-stack and link overheads.


\begin{table}[t]
  \setlength{\belowcaptionskip}{-10pt}
\centering
\footnotesize
\resizebox{\columnwidth}{!}{%
\begin{tabular}{@{}lllll@{}}
\toprule
 & HBM & HBF & CXL-flash & NVMe SSD \\
\midrule
Interface & pseudo-ch & UCIe/AXI & CXL.mem & PCIe/NVMe \\
Granularity & 32--64B & 64B/4KiB & 64B & 512B--4KiB \\
Latency & 100\,ns & 1--15\,$\mu$s & $\mu$s--ms & 100\,$\mu$s \\
Device GC & no & \textbf{no} & partial & yes \\
Capacity & 80--192\,GB & configurable 512\,GiB span & TB-class & TB-class \\
BW & 1--8\,TB/s & nominal 3.07\,TB/s & 10s GB/s & 3--14\,GB/s \\
\bottomrule
\end{tabular}
}%
\caption{\textbf{Comparison of representative memory and storage tiers.}
Parameters are based on respective industry specifications (JEDEC HBM3/3e, OCP HBF v0.7.0~\cite{ocp_hbf_v07}, CXL 3.1~\cite{cxl}) and enterprise SSD datasheets\cite{samsung983dct,micron9550}. HBF latency reflects representative flash media access and buffer-hit timing profiles.
}
\label{tab:tiers}
\end{table}


\begin{table*}[htbp]
\centering
\footnotesize
\setlength{\tabcolsep}{3pt}
\renewcommand{\arraystretch}{1.25}
\begin{tabular}{@{}lcccc@{}}
\toprule
Platform & \shortstack{Configurable\\GPU timing} & \shortstack{NAND page\\timing} & \shortstack{Load/store\\flash access} & \shortstack{HBF request\\and completion} \\
\midrule
GPGPU-Sim / Accel-Sim~\cite{gpgpusim,accelsim} & Yes & No & No & No \\
MQSim~\cite{mqsim} & No & Yes & No & No \\
SimpleSSD (gem5 CPU)~\cite{simplessd, gem5} & No & Yes & No & No \\
FEMU (BlackBox)~\cite{femu} & No & Yes & No & No \\
Cylon~\cite{cylon} & No & Yes & Yes & No \\
\textbf{HBF-Sim (ours)} & \textbf{Yes} & \textbf{Yes} & \textbf{Yes} & \textbf{Yes} \\
\bottomrule
\end{tabular}
\caption{\textbf{Platform capabilities for GPU--HBF studies.} Yes/No indicates support in the listed configurations.}
\label{tab:platforms}
\vspace{0pt}
\end{table*}

\subsection{HBF in the GPU Memory Hierarchy}


HBF becomes a new package-local flash tier alongside HBM on GPU, as shown in
Figure~\ref{fig:hardware}. Its NAND dies and logic/base die connect to the GPU
through package-level links, but this physical proximity does not give HBF
DRAM-like access behavior. As Table~\ref{tab:tiers} summarizes, HBF combines a
memory-style interface with page-buffered reads, channel-affine media resources,
and non-posted page writes. Modeling its place in the GPU memory hierarchy
therefore requires both GPU execution timing and these flash semantics.

Accel-Sim and GPGPU-Sim already model instruction execution, caches,
interconnects, and warp timing~\cite{accelsim,gpgpusim}. HBF-Sim extends this
path with HBF page, channel, and completion state while retaining the ordinary
DRAM path. GPU cache structures and HBF page-tracking structures have distinct
roles; the HBF cache-routing policy is described in Section~\ref{sec:design}.



Consequently, studying HBF in this hierarchy requires a \textbf{closed timing loop}: GPU execution generates memory requests, HBF queues and services them under page and channel constraints, and their completion timing dictates when dependent GPU instructions can proceed. Capturing this bidirectional feedback is essential for understanding HBF performance, exposing critical limitations in existing simulation tools (\S\ref{sec:simulator_limits}).

\section{Motivation}
\label{sec:motivation}

\subsection{Limitations of Existing Simulators}
\label{sec:simulator_limits}

Table~\ref{tab:platforms} compares the capabilities of related works about GPU--HBF architectures. GPU timing covers instruction execution, caches, and request
issue. NAND timing covers page reads and programs, while load/store flash
access provides a memory-style interface. HBF simulation must connect these
capabilities so that page service affects GPU execution, including write
acknowledgements after programming.

GPGPU-Sim and Accel-Sim~\cite{gpgpusim,accelsim} model GPU execution and memory
timing, but their base configurations lack NAND page operations. Increasing
DRAM capacity and latency does not add page-buffer reuse or page-level
coalescing. MQSim~\cite{mqsim}, SimpleSSD~\cite{simplessd}, and FEMU~\cite{femu}
provide flash-device timing. SimpleSSD integrates with a simulated CPU,
while FEMU emulates a device for guest software. Cylon~\cite{cylon} supports
load/store access to CXL-attached flash. The listed configurations do not
connect simulated GPU requests to HBF page service and return completion
dependencies to GPU execution.


A controller policy can change both the time needed to serve requests and
when subsequent GPU requests reach HBF. For example, merging reads to the
same page reduces the number of outstanding page operations. Under admission
pressure, this can free controller capacity sooner and let waiting GPU
requests proceed earlier. A replay with fixed arrival times keeps the
original request schedule and misses this feedback, potentially
underestimating the benefit of merging. Evaluating such policies therefore
requires GPU request timing to respond to HBF service. Section~\ref{sec:mshr}
compares fixed-arrival replay with coupled GPU execution to demonstrate this
effect.

\textbf{In summary,}
A GPU--HBF simulator needs both page-level media behavior and completion
feedback to GPU execution. This requirement motivates the mapping, read,
write, and tracing mechanisms in Section~\ref{sec:design}.

\subsection{Various R/W Patterns of HBF}

HBF must serve different access patterns within the same system.
Table~\ref{tab:wls} summarizes representative data classes in LLM and AI
workloads. Sequential weight reads require sustained data delivery. Repeated
accesses to KV data, such as shared prefixes~\cite{hydragen}, make page reuse relevant, while appending KV data requires
write aggregation and completion tracking. Scattered RAG-vector accesses can
fetch a full NAND page for a small useful region, increasing read amplification.

\begin{table}[t]
\centering \footnotesize
\begin{tabular}{@{}llll@{}}
\toprule
class & pattern & size & modeling focus \\
\midrule
weights & seq., read-only & 100s GB & bandwidth \\
precomputed KV & reuse reads & 100s GB--TB & page reuse \\
ephemeral KV & append-writes & 10s GB & write ordering \\
RAG vectors & random gathers & 100s GB & read amplification \\
\bottomrule
\end{tabular}
\caption{\textbf{LLM access patterns and HBF modeling focuses.}
These categories guide the configuration of our evaluation workloads.}
\label{tab:wls}
\end{table}

These patterns place different demands on the controller. Requests to the
same page can share a page read or reuse buffered data. Requests to different
pages need enough independent media resources to execute in parallel.
When reads and writes overlap, a long page program can delay reads that need
the same resources. A policy that helps one access pattern may therefore
provide little benefit for another. Studying these effects requires control
over address mapping, request coalescing, and media scheduling.

We use controlled request streams to vary concurrency, locality, and write
pressure separately. Keeping the request and completion semantics consistent
allows us to compare policies under the same workload. Request traces connect
GPU accesses to queueing, page service, and completion, helping explain whether
a change reduces NAND work or shortens GPU execution. 
These representative categories guide the synthesis of our microbenchmarks, allowing us to isolate and study specific architectural mechanisms under controlled traffic conditions.

\textbf{In summary,}
A GPU--HBF simulator needs configurable mapping, read, and write policies
under a common timing model. This requirement motivates an extensible design
that supports different access patterns and exposes their effects on both
media activity and GPU execution.

\section{HBF-Sim Simulator Design}
\label{sec:design}
\begin{figure*}[t]

\setlength{\abovecaptionskip}{-20pt}
\setlength{\belowcaptionskip}{-15pt}
\centering
\includegraphics[width=0.95\textwidth]{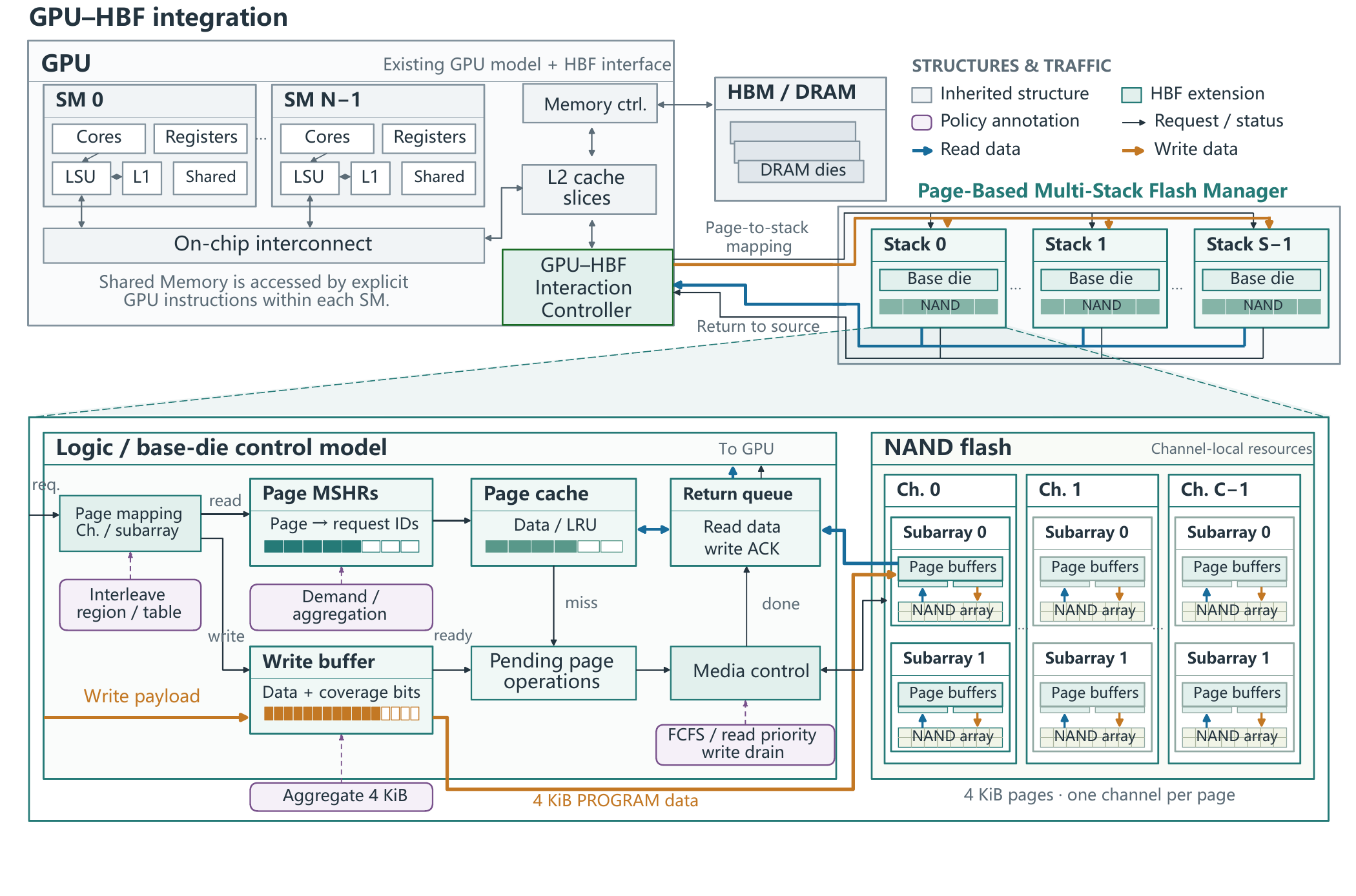}
\caption{\textbf{HBF-Sim architecture.}  The upper part shows GPU memory paths and independent HBF stacks. The lower part expands one stack into its control structures and channel-owned NAND resources. Rectangles denote modeled structures, while the figure's policy annotations identify configurable methods. Thin arrows carry requests or control/status. Thick blue and orange arrows carry read and write data.}
\label{fig:arch}
\end{figure*}

HBF-Sim models two cooperating layers that determine HBF behavior:
(i) the \emph{GPU--HBF interaction controller}, which sits above the base die
and governs the load/store contract between the GPU and HBF, and (ii) the
\emph{page-based parallel stack storage}, which organizes data placement,
coalescing, and media access inside the HBF stacks. The former owns granularity
alignment, address mapping, memory management, and flash management. The latter
owns coalescing-based data storage and data retrieval.
Figure~\ref{fig:arch} illustrates these functions in the GPU--HBF organization.
Its upper view shows the GPU memory paths and the connections to independent
HBF stacks. The expanded stack below separates request-tracking structures
from data buffers and channel-owned NAND resources. Purple annotations
identify the mapping, aggregation, and scheduling choices implemented on
those structures. 

HBF-Sim places an explicit boundary between GPU timing and HBF media timing.
The GPU produces sector or cacheline requests and consumes completion events
through load/store interfaces; the HBF system owns page, channel, and
media state. GPU requests, HBF pages, and NAND resources are separated so that
GPU timing, interface queueing, and media service remain distinguishable.

\subsection{GPU--HBF Interaction Controller}
\label{sec:interaction}
The interaction controller translates GPU accesses into HBF page operations. It identifies the page containing each request and assigns the
request to a stack, channel, and subarray. Each request retains its source GPU partition so that the completion
response returns to the correct requester.

\subsubsection{Cacheline-to-page Granularity Alignment}
GPU requests must be aligned to the page granularity used by HBF. The GPU emits
32\,B sectors within 64 or 128\,B cache-line transactions, while HBF reads
and programs 4\,KiB pages. The controller groups requests by page so the read
and write modules can combine accesses to the same page. Let $B$ be the start
of the configured HBF address range, $a$ a request address in that range, and
$P=4096$\,B the page size. We compute the zero-based logical page number as

\[
p(a)=\left\lfloor\frac{a-B}{P}\right\rfloor.
\]
Thus, $a-B$ is a byte offset within the HBF range. For example, accesses at
$B+64$ and $B+128$ both belong to page~0, while $B+4096$ belongs to page~1.

\subsubsection{Channel-oriented Address Mapping}
A logical HBF system contains one or more independent stacks. Each stack exposes
1, 2, 4, 8, or 16 channels in the standard profiles. Configurations beyond 16 channels are supported as experimental extensions. GPU memory partitions do not create
additional copies of a stack or multiply its bandwidth. The source subpartition
is retained for return routing only. Total capacity is divided evenly into
erase-block-aligned stack-local capacities. Page interleave or contiguous
capacity slices select a stack. Stack-local addresses are restored before
completion returns to the GPU. After admission, each request is assigned to one
stack, one channel, and its fixed subarray resources. There is no cross-channel
resource sharing.

We implement three placement policies. Page interleave distributes HBF-relative pages
across channels. Contiguous placement assigns regions to selected channels. An
explicit mapping table can place data classes such as weights and KV pages. The
grouped placement evaluated in Section~\ref{sec:locality} uses this hook to
rotate after every two adjacent pages. The mapping uses the address relative to
the configured HBF base address. An abstract per-channel bandwidth credit models
the host link. We limit outstanding operations per channel to model queue backpressure.
The host-link abstraction is discussed in Section~\ref{sec:disc}.

\subsection{Page-Based Multi-Stack Flash Manager}
\label{sec:flashmanager}
We design the flash manager to use page-level parallelism across the channels
and subarrays of each stack. The write module collects GPU stores into full
pages. The read module combines requests to the same page and reuses buffered
data when ordering permits. Both modules operate on 4\,KiB NAND pages.

\subsubsection{Merge-based Write Module}

We use a per-page write buffer to combine GPU stores into full-page
programs. The buffer tracks both the data and the byte positions covered
by each store. A page is flushed when it is full or when buffer pressure
requires space. It is also flushed when the configured deadline expires, when a
same-page read requires ordering, or during an idle tail drain. The default
\texttt{strict} timeout policy records an incomplete-page error and does not
label the operation as a full-page program. The opt-in \texttt{partial}
compatibility policy pads and programs a partial page while recording both the
partial program and padding bytes. The policy is selected with
\texttt{hbf\_write\_timeout\_policy}. Flushes are serialized per page. While a
page MSHR is in flight, later writes remain buffered and cannot overwrite the
entry. Erase-before-write and sequential in-block allocation follow the HBF
profile. The controller returns the write response only after programming completes. This
non-posted completion propagates write backpressure to the GPU store path.

\begin{figure}[tb]
\centering
\setlength{\abovecaptionskip}{-5pt}
 \setlength{\belowcaptionskip}{-15pt}
\includegraphics[trim=0cm 0cm 0cm 6cm, clip, width=0.85\columnwidth]{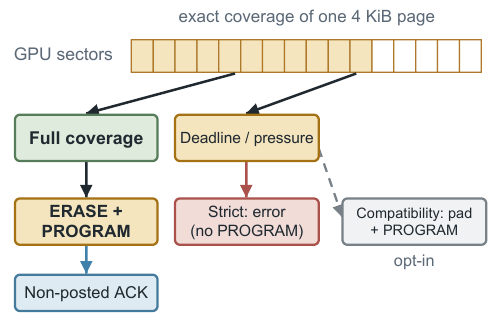}
\caption{\textbf{Page-based write aggregation.}
Incomplete pages trigger an error on flush or are padded in compatibility mode.
The GPU receives a completion response only after the full page is written to NAND.}
\label{fig:details}

\end{figure}

We use the union of written byte positions to determine whether a page is
complete (Figure~\ref{fig:details}). Rewriting
an already covered region does not bring an incomplete page closer to full
coverage. A full page can be programmed when its resources are available. An incomplete page forced to flush follows the strict-error or opt-in padding
path. The acknowledgement follows PROGRAM because a GPU store completes only
after the data has been programmed.

\subsubsection{Page-grained Read Module}

We use page-keyed MSHRs to combine outstanding reads to the same page.
Sector or cache-line requests to one 4\,KiB page can share one NAND read. In \emph{demand mode}, the
first request dispatches as soon as the page is admitted. Later requests to that
page merge with the in-flight MSHR. In \emph{aggregation mode}, the controller
waits for a configured time window or byte threshold before issuing the page
read. The aggregation wait is bounded by that configured window. Aggregation
mode can reduce duplicate page reads for dense same-page traffic. Aggregation is an optional policy. The interface exposes these choices as
\texttt{hbf\_read\_mode}, \texttt{hbf\_read\_agg\_window}, and
\texttt{hbf\_read\_agg\_threshold}. The default aggregation threshold is one
4\,KiB page.

Each read is scheduled on a subarray owned by the page's channel.
The scheduler checks subarray availability and a configurable limit on active
operations before issuing a page operation. 
Each sub-array implements READ, PROGRAM, and ERASE states with configurable timings.
The default profile uses $t_R=15\,\mu$s, $t_{\mathit{PROG}}=200\,\mu$s, and
$t_{\mathit{BERS}}=2\,ms$. Each sub-array has two NAND page buffers. 
A buffered-page reread pays the buffer-hit latency instead of a full media read when ordering permits. 
A logic-die page cache provides LRU reuse across kernels and access windows.

\subsection{GPU--HBF Request Lifecycle}

\begin{figure}[!bt]
\setlength{\abovecaptionskip}{0pt}
\setlength{\belowcaptionskip}{0pt}
\centering
\includegraphics[width=0.95\columnwidth,keepaspectratio]{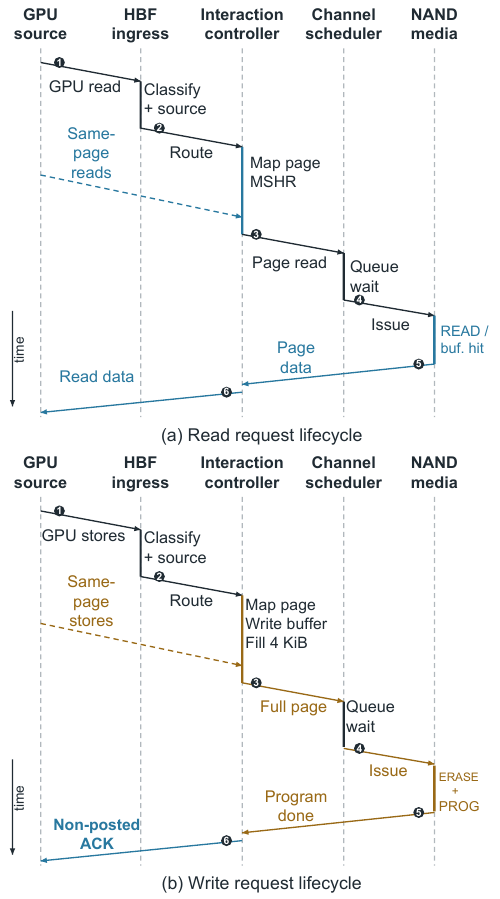}
\caption{\textbf{Request lifecycle through HBF-Sim.} A GPU read or write request enters
the interaction controller, is mapped to one stack and channel, and follows
page tracking, queueing, and NAND service before read data or write status
returns to the source GPU partition.}
\label{fig:lifecycle}
\end{figure}

We connect these modules through the read and write sequences shown in
Figure~\ref{fig:lifecycle}. Time advances downward in the diagram. The vertical spacing shows event
order rather than measured durations.
In panel (a), the initial read is mapped to a stack and tracked by page, so
later same-page requests can join the outstanding operation. Channel waiting
precedes READ or page-buffer service, after which data returns to the source.
Panel (b) shows the additional write dependency: same-page stores first build
complete coverage, then the page waits for service and is programmed, with
erasure if needed. The write acknowledgement is returned only afterward.

Each request retains its source subpartition for return routing. The cube advances its controller, channel, and NAND state
machines on the DRAM clock. Completed requests enter the source partition's
return path and then follow the existing interconnect. Cache fills are handled
by the ordinary L2 fill path when a corresponding cache miss is pending.
DRAM requests continue through the existing DRAM path.

\ddparagraph{Clock domains and cost.}
The controller advances HBF media timers ($t_R$, $t_{\mathit{PROG}}$,
$t_{\mathit{BERS}}$, page-buffer hits, and page-cache hits) with the DRAM clock
used by the simulator. GPU-visible latency is converted to GPU core cycles when
a request completes. Statistics retain both units and report the DRAM frequency
and tick-to-$\mu$s conversion. The controller keeps one page-keyed MSHR map, one
pending-page FIFO, one write-buffer map, and a set of busy subarrays. Only busy
sub-arrays are advanced on each tick. The per-tick cost therefore follows the
number of in-flight operations rather than the full array size. Scheduling scans
the pending queue within the outstanding-request cap.

\subsection{Trace Generation}
\label{sec:trace}

We implement request tracing to connect GPU accesses with controller
decisions and media operations. Each record identifies the request state
and the work performed to serve it.

\ddparagraph{State and accounting.}
We record each request's identifier, operation, address, size, source, and
arrival and completion times. Each request advances through \textsc{Ingress}, \textsc{Mapped},
optional write \textsc{Buffered}, \textsc{Queued}, \textsc{MediaOp},
\textsc{Return}, and \textsc{Completed} states. GPU L2 and CUDA Shared Memory
remain distinct from the HBF page cache and NAND page buffers. Bounded queues
may delay a request but cannot change its owner. Device timers advance in
DRAM-clock ticks. GPU-visible latency is recorded in core cycles. Read
amplification is $\mathit{media\ bytes}/\mathit{requested\ bytes}$, while write
accounting records coverage, programs, padding, and ordering errors.

We expose configuration parameters for per-channel bandwidth, NAND timings, channel count
and mapping, subarray count, page-cache and page-buffer sizes, write-buffer
depth and deadline, placement, and media mode. Placement can be contiguous,
page-interleaved, or selected by an explicit mapping table. Statistics include
per-channel request and operation counts and link-stall ticks. They include
first-request latency, aggregation wait, page-read count, requested bytes, media
bytes, read amplification, and MSHR hits. They also include page-buffer and
page-cache hits, partial-page programs, timeout flushes, idle drains, padding or
strict-policy errors, read-after-write ordering, capacity utilization, and
per-zone PEC spread.

The optional CSV trace records \texttt{sim\_cycle}, \texttt{request\_id},
\texttt{source\_subpartition}, \texttt{op}, \texttt{address}, \texttt{page},
\texttt{channel}, \texttt{subarray}, \texttt{state}, and \texttt{bytes}. It also
records \texttt{queue\_depth}, \texttt{latency}, \texttt{cache\_hit},
\texttt{mshr\_hit}, and \texttt{error}. The trace file and trace detail are
selected with \texttt{hbf\_trace\_file} and \texttt{hbf\_trace\_level}. The
ordered 15-column schema is frozen as \texttt{hbf-trace-v1}. Validation rejects
missing, added, or reordered columns. Full-system traces preserve GPU injection
effects. Device-only replay can inject the same request stream without the GPU
issue-rate limit.

\subsection{Implementation}
\label{sec:implementation}

We implement HBF-Sim in C++ as an Accel-Sim extension and maintain the
HBF source files in \texttt{hbf/}. The setup script applies the
integration patch to a pinned simulator revision and copies the active cube,
controller, media, and trace sources into the build tree. This keeps generated
simulator files separate from the maintained HBF module.

We instantiate one logical HBF system with the configured number of
independent stack controllers. Each stack owns its channels, controller queues,
page cache, FTL, and trace writer. The system layer owns address routing and
fair return arbitration. The controller stores page-keyed MSHRs and per-page
write buffers. It instantiates concrete subarray state lazily, advances only the
active-subarray set, and skips the HBF tick when the complete system is idle.
The same sources build through the Make and CMake paths.

The artifact scripts compile the validation kernels, create timestamped
configurations, and store binary and trace hashes beside each run. The trace
remapper changes only global addresses and writes a manifest with the input and
output hashes. The device replay tool consumes the same request schema as the
simulator and is used to separate HBF service from GPU injection.
%



\section{Platform Validation and Evaluation}
\label{sec:evaluation}

We evaluate request handling, timing feedback, page service, and resource
scaling. We first check functional behavior and compare fixed-arrival replay
with coupled GPU execution. We then cross-check page service with MQSim,
measure media scaling, and run a Qwen3 decode trace. Finally, we measure
simulator cost and addressable capacity. Section~\ref{sec:cases} uses the same
platform to evaluate placement and read/write isolation policies.

\subsection{Methodology and Functional Validation}
\label{sec:eval_setup}
\label{sec:validation}

Unless otherwise specified, we use Accel-Sim with a 1.132\,GHz GPU core
clock and an 850\,MHz DRAM clock. Scaling retains an 80-SM configuration. The placement and mixed-write
studies use 8 shader clusters and 8 memory partitions. 
Default READ/PROGRAM/ERASE timings are 15/200/2000\,$\mu$s. The placement diagnostic uses 20/40/80 DRAM ticks.
These are model parameters rather than hardware measurements. Raw configurations, traces and binary hashes are retained.

We measure request latency from request creation to completion return by
the HBF controller, including waiting before controller acceptance.
We report the 95th percentile (p95) of read and write request
latencies in the read/write interference study.
GPU execution time covers the entire workload, including write completion.
Both request latency and GPU execution time are measured in GPU clock cycles.

Our routing tests preserve all 64 request/completion pairs at 1--16 channels.
The shared-staging probe produces 32 HBF requests, one page read and 31 MSHR
hits. Demand and aggregation finish in 5,508 and 5,517 cycles. A remapped LUD
trace preserves all 158 request/completion pairs. C++ coverage tests check
repeated fragments and page boundaries. Successful strict-write traces require
complete coverage before PROGRAM and acknowledgements after programming.
These checks exercise different parts of the request path. Pairing request IDs
with completion IDs detects lost or duplicate returns. The shared-staging
probe checks that one page operation can serve multiple waiting requests.
For writes, repeated fragments must not count as new byte coverage, and a
complete page must be programmed before its acknowledgement is released.
Our tests validate request handling in these workloads. Hardware calibration
remains outside this evaluation.

\subsection{MSHR Benefits and Timing Feedback}
\label{sec:mshr}

We test whether a fixed device-input trace preserves the benefit of page-keyed
MSHR merging when admission pressure changes. The closed-loop configuration
lets GPU execution and upstream queues respond to HBF completions. The
open-loop configuration freezes controller-ingress times from a source run.
Both use the same C++ HBF controller, media model, admission limits and return
credits. The replay driver retains source request order and enforces at most one
admission per memory partition per DRAM tick, but does not execute GPU cores.

We run 24 GPU configurations: 16 pages, 1 or 32 distinct 128\,B entries per
page, 32/128/4,096 admission slots, 1/15\,$\mu$s reads, and merging on/off.
Eight SMs and eight memory partitions use 1.132/0.850\,GHz core/DRAM clocks. Four HBF channels, 32 subarrays and eight active slots are fixed.
The logic-die cache is disabled and NAND page buffers are enabled.
Every source replay reproduces its complete controller event trace exactly.
We then replay each source into the opposite merging configuration, yielding
24 counterfactual runs with identical address/size multisets and conserved
request/completion pairs. 

\begin{figure}[!tp]
\centering
\setlength{\abovecaptionskip}{-5pt}
 \setlength{\belowcaptionskip}{-5pt}
\includegraphics[width=\columnwidth]{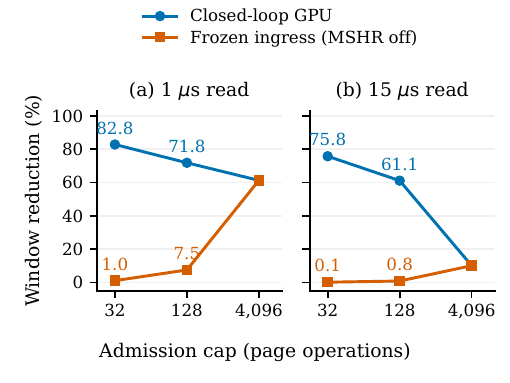}
\caption{\textbf{Frozen ingress understates MSHR benefits under admission pressure.}
HBF completion-window reduction relative to merging disabled, using
16 pages and 32 entries per page. Replay fixes the baseline's ingress times.}
\label{fig:locality}
\end{figure}

We compare the window from first HBF admission to last HBF completion.
This interval is available in both GPU simulation and device-only replay.
Our evaluation shows that fixed-arrival replay understates the benefit of
merging under admission pressure (Figure~\ref{fig:locality}). We use the input
trace captured with merging disabled as the replay baseline. At 15\,$\mu$s and 128 slots, the closed
loop reduces this window from 87,703 to 34,101 cycles (61.1\%). Frozen ingress
instead gives 86,987 cycles (0.8\% reduction), a 60.3-percentage-point
difference in predicted benefit. The full GPU kernel correspondingly falls
from 93,699 to 40,164 cycles. With 32 slots, the window reductions are
75.8\% and 0.1\%. With 4,096 slots, both predict 10.0\%.
The 1\,$\mu$s controls exhibit the same pressure-dependent pattern.
All sparse-input controls and reverse transfers are retained in the artifact.

Our traces explain this difference through request admission. At 128 slots, enabling
merging contracts the closed-loop ingress span from 86,972 to 414 cycles:
fewer distinct page operations occupy admission slots, allowing upstream GPU
requests to advance. Frozen ingress retains the source run's waiting even
after merging is enabled. Conversely, disabling merging while replaying the
merging-enabled trace changes the target window by only 1.8\% at this point. The abstraction difference is directional. At 4,096 slots the source ingress
schedules coincide and replay agrees. Our comparison shows that admission feedback affects the estimated benefit
of controller policies under pressure. 

\subsection{External Page-Service Cross-Check}
\label{sec:external_compare}

We compare the HBF device driver with unmodified MQSim~\cite{mqsim}
(revision \texttt{51f0f2d}), restricting the comparison to their common
read-page service. Eight paired configurations use 64 unique 4\,KiB pages,
one or four channels with one serial die/subarray per channel, and
1/15\,$\mu$s reads. Page arrivals are either 50\,$\mu$s apart or a burst
spaced by 10\,ns. Each logical page is one eight-sector NVMe request in
MQSim and 32 merged 128\,B requests in HBF-Sim. Both service 256\,KiB.
HBF-Sim performs exactly 64 page operations. Data/page caches are disabled.
MQSim uses ideal address mapping and no preconditioning. The read-only run
avoids GC and write-management differences.

Nominal channel bandwidth is 192\,GB/s in both models. MQSim uses a synthetic
192-byte channel at 1,000\,MT/s and four 192\,GB/s PCIe lanes to keep host
bandwidth from dominating. These are comparison parameters, not a projected
SSD. Native NVMe/ONFI processing remains in MQSim. HBF-Sim retains its
bandwidth-credit link and GPU-sized fragments. We therefore compare logical
arrival-to-completion windows, without equating the two device interfaces.

\begin{table}[t]
\centering\footnotesize
\setlength{\abovecaptionskip}{3pt}
\setlength{\belowcaptionskip}{-5pt}
\begin{tabular}{@{}rrrrr@{}}
\toprule
Channels & $t_R$ ($\mu$s) & HBF-Sim & MQSim & Difference \\
\midrule
1 & 1 & 64.032 & 85.126 & +32.9\% \\
4 & 1 & 16.080 & 21.316 & +32.6\% \\
1 & 15 & 960.032 & 981.126 & +2.2\% \\
4 & 15 & 240.080 & 245.316 & +2.2\% \\
\bottomrule
\end{tabular}
\caption{\textbf{External read-service comparison.} Burst completion windows
in $\mu$s. Difference is $(T_{\mathrm{MQSim}}/T_{\mathrm{HBF}}-1)$.
The models retain their respective interface overheads. These differences
are not errors against HBF silicon.}
\label{tab:mqsim_comparison}
\end{table}

\begin{figure*}[!tp]
\centering
\setlength{\abovecaptionskip}{-5pt}
 \setlength{\belowcaptionskip}{-10pt}
\includegraphics[width=0.98\textwidth]{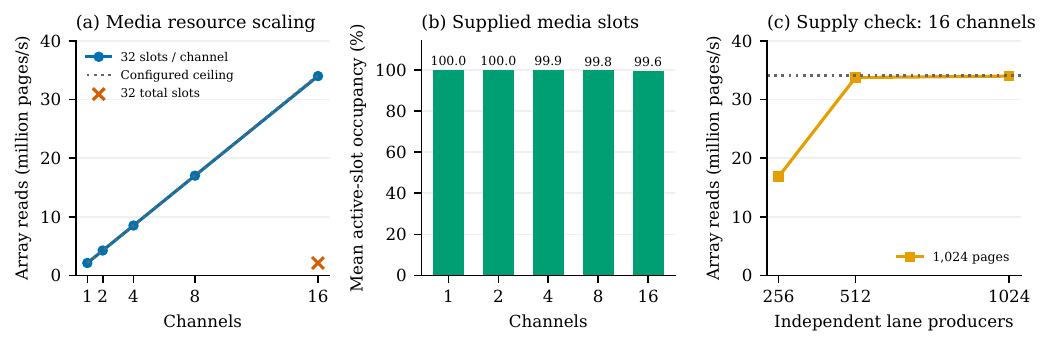}
\caption{\textbf{Media scaling with 32 subarrays and active slots per channel at 15\,$\mu$s reads.} Throughput counts array pages, not GPU payload bytes. Occupancy and producer-count controls check saturation.}
\label{fig:scaling}
\end{figure*}

Our comparison shows similar scaling in both models, with an offset caused
by their different interface models (Table~\ref{tab:mqsim_comparison}). Increasing $t_R$ from 1 to 15\,$\mu$s adds
896\,$\mu$s with one channel and 224\,$\mu$s with four in both models,
matching $64\Delta t_R/C$. MQSim retains, for example, a 290\,ns single-page
ONFI command phase in this configuration. Its burst windows exceed HBF-Sim's
by 2.2\% at 15\,$\mu$s and about 33\% at 1\,$\mu$s. The isolated controls
also retain an interface offset. Their long arrival gaps are excluded from
claims of close agreement. This external check supports the implemented
read-timing and parallel-service trends while exposing the scope of our link
abstraction. It does not calibrate HBF silicon, GPU timing, or write behavior.

\subsection{Media Resource Scaling}
\label{sec:scaling}

We test media scaling with 1,024 lanes, each reading a different page
to provide independent requests. Each 4\,B load generates a 32\,B sector request
and one 4\,KiB array read. Both cache shortcuts are disabled. We retain GPU
backpressure and scale from 1 to 16 channels, with 32 subarrays and active
slots per channel, 4,096 admission entries and default media timings.

Our evaluation shows that array-service bandwidth increases from 8.738
to 139.285\,GB/s (15.94$\times$), with 99.6--99.99\% media occupancy
(Figure~\ref{fig:scaling}).
At sixteen channels, 256 producers fill only half of the 512 slots.
512/1,024 producers deliver 138.291/139.285\,GB/s. Doubling the stream to
2,048 pages/producers changes throughput by about 0.2\%, while fixing total
slots at 32 restores 8.738\,GB/s. Array bandwidth counts internal page reads:
the corresponding sixteen-channel GPU sector bandwidth is only 1.088\,GB/s.
Our fixed-slot control shows that the gain comes from additional media
parallelism. Increasing the channel count alone does not improve throughput
when the total number of active slots remains fixed. The producer-count controls check that the main sweep has
enough outstanding work to occupy the added resources. 

\subsection{LLM Decode Workload}
\label{sec:llm}

We use Qwen3-1.7B~\cite{qwen3_1p7b} to exercise the GPU--HBF path with a
real pretrained language model. The standard Transformers implementation
runs in BF16 with eager attention, batch size one, and thinking disabled.
The model has 28 layers, hidden dimension 2,048, 16 query heads and eight KV
heads. Native inference on an RTX 4090 processes a 27-token prompt and
produces eight tokens using greedy decoding. We use Accel-Sim's NVBit-based
tracer~\cite{nvbit} to capture the second decode
forward pass after prefill, with a 29-token context, preserving all 1,819
GPU kernels across the full model. The native run supplies the input token
and KV state for this captured step.

The trace remapper translates only the recorded parameter allocations into
the HBF address range, preserving offsets and shared-weight aliases. KV,
activations and library scratch remain on the ordinary GPU-memory path.
Replay uses an SM80 timing profile with eight SMs, eight memory partitions,
and 1.410/1.512\,GHz core/DRAM clocks. One HBF stack provides 16 channels,
512 total subarrays, 512 active-operation slots and 4,096 admission entries.
READ/PROGRAM/ERASE timings are 15/200/2000\,$\mu$s. MSHR merging and NAND
page buffers are enabled. The logic-die page cache is disabled. 
This configuration is evaluated separately from the mechanism studies.

Our simulator completes all 1,819 captured kernels. We report the final
execution and traffic counters in Table~\ref{tab:qwen_decode}.
The sequence takes 352,434,530 GPU cycles (249.95\,ms). The controller
records 107,540,992 read requests and 107,540,992 read completions.

\begin{table}[t]
\centering\footnotesize
\setlength{\abovecaptionskip}{5pt}
 \setlength{\belowcaptionskip}{-10pt}
\begin{tabular}{@{}ll@{}}
\toprule
Metric & Result \\
\midrule
Captured / completed GPU kernels & 1,819 / 1,819 \\
GPU kernel-sequence time (ms) & 249.95 \\
HBF read requests & 107,540,992 \\
HBF controller read completions & 107,540,992 \\
Returned sector bytes & 3.441 GB \\
Array-read bytes & 3.441 GB \\
\bottomrule
\end{tabular}
\caption{\textbf{Qwen3-1.7B decode execution and HBF traffic.} Values are taken from final simulator counters. Time covers the GPU kernel sequence. Array bytes are page services minus NAND page-buffer hits, multiplied by 4,096 bytes. GB denotes $10^9$ bytes.}
\label{tab:qwen_decode}
\end{table}

This experiment demonstrates HBF-Sim's capability to execute the complete GPU-kernel sequence of a captured Qwen3-1.7B decode step. Starting from cold simulated caches, the reported execution time focuses directly on the memory-bandwidth-bound GPU kernel execution during active token generation, with model weights residing in HBF while KV caches and activations remain in GPU memory.

\subsection{Simulator Cost and Capacity}

We measure simulator cost with five alternating idle-path repetitions per mode. 
The median host times are 6.768/6.964\,s and peak RSS of 58,832/59,016\,KiB with HBF disabled/enabled.
Both execute 181,707 cumulative cycles over 32 kernels. 
We also test a four-stack, 512\,GiB address space with a sparse probe. 
It completes 20 sectors across five boundary/interior pages without mapping
errors or eager metadata allocation. It tests addressing and capacity accounting,
not full-capacity sustained performance.

\section{HBF-Sim-based Optimization Cases}
\label{sec:cases}
Having checked the mechanisms in Section~\ref{sec:evaluation}, we use
HBF-Sim to study two policy questions: how page placement affects read service
and execution, and when resource isolation helps protect reads from writes.
These cases compare static policies through the mapping and scheduling
interfaces while retaining the GPU request/completion path. They use controlled
GPU workloads. Their findings concern the tested access patterns and resource
configurations.

\subsection{Locality/Parallelism-Aware Placement}
\label{sec:locality}

Page placement can reduce repeated page service and lower read-side
page-service amplification. We first measure this benefit and its effect on
kernel execution, then examine how placement affects subarray parallelism.

\begin{figure}[t]
\centering
\setlength{\abovecaptionskip}{0pt}
\setlength{\belowcaptionskip}{0pt}
\includegraphics[trim=0cm 7cm 0cm 0cm, clip, width=\columnwidth]{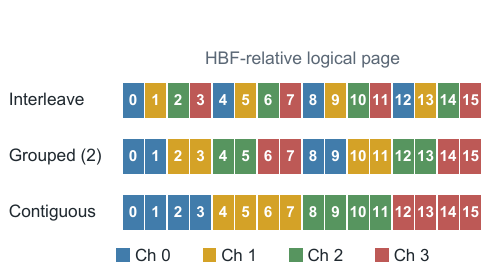}
\caption{\textbf{Channel assignments under three placement policies.}
Interleave rotates channels after each page; grouped placement rotates after
every two pages. Contiguous placement assigns consecutive active pages to each channel.}\vspace{-10pt}
\label{fig:case_mapping}
\end{figure}

\ddparagraph{Policies and setup.}
We compare page interleave, two-page grouping, and contiguous placement
through the address-mapping interface in Section~\ref{sec:design}.
Figure~\ref{fig:case_mapping} shows their channel assignments.
Interleave rotates channels after each page, while grouped placement rotates
after every two pages. Contiguous placement divides the configured capacity
into channel-owned regions. Each mapping determines both the channel and
subarray of a page and remains fixed throughout a run.

The study covers nine configurations: three mappings and 1/8/32 distinct
128\,B entries per page, with shortened read timing ($t_R=20$ DRAM ticks).
We use a CUDA read microbenchmark with 16 pages, four channels, 32 total
subarrays, eight active slots, and an admission cap of 256.
MSHR merging and NAND page buffers are enabled; the logic-die page cache
is disabled.

\begin{figure}[t]
\centering
\setlength{\abovecaptionskip}{-5pt}
\setlength{\belowcaptionskip}{-5pt}
\includegraphics[width=\columnwidth]{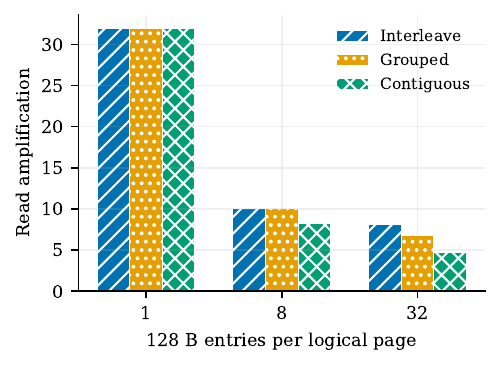}
\caption{\textbf{Page-service amplification under shortened timing ($t_R=20$ DRAM
ticks).} At 32 entries per page, contiguous placement reduces page-service
amplification by 41.9\% relative to interleave. Page services include
NAND page-buffer hits.}
\label{fig:case_service}
\end{figure}

\ddparagraph{Reducing page-service amplification.}
With $B$ denoting requested read bytes, page-service amplification and
array-read amplification are:
\[
A_s=\frac{4096N_{\mathrm{service}}}{B},\qquad
A_a=\frac{4096N_{\mathrm{array}}}{B}.
\]
$N_{\mathrm{service}}$ counts page services, including page-buffer hits,
while $N_{\mathrm{array}}$ counts NAND array reads.
Figure~\ref{fig:case_service} shows that placement reduces page-service
amplification as accesses become denser within each page.
At 32 entries per page, interleave, grouped, and contiguous placement require
129, 108, and 75 page services, respectively. Contiguous placement lowers
amplification from 8.0625 to 4.6875 relative to interleave, a 41.9\% reduction.

\ddparagraph{Effect on kernel execution.}
All three mappings perform 16 array reads at 32 entries per page; the
remaining page services are served by the NAND page buffers. Thus, the
reduction in page-service amplification leaves array work unchanged.
Kernel time rises slightly from 6,505 cycles with interleave to 6,550 cycles
with contiguous placement, an increase of 0.69\%.
The next experiment examines how the distribution of accesses across
subarrays affects execution time.

\ddparagraph{Resource conflicts and parallelism.}
We use a separate 36-run study with default read timing ($t_R=15\,\mu$s),
64/128 active pages, 1/8 entries per page, and logical-page strides of 1/4/32.
We compare built-in interleave, an explicit modulo-channel map, and
active-page striping. For active pages $p_i=iD$, where $D$ is the page stride
and $C$ is the channel count, explicit modulo assigns channel $p_i\bmod C$
and active-page striping assigns $i\bmod C$.
The resource budget remains four channels, 32 total subarrays, and eight
active slots, with an admission cap of 4,096.
The two studies use different simulator binaries; policy comparisons are
made within each study.

The mappings also differ in their subarray assignments. For logical page $p$,
built-in interleave uses $\lfloor p/C\rfloor$ as the channel-local page index,
while explicit maps use $p$. The local subarray index is this value modulo
the number of subarrays in the channel. The explicit modulo map therefore
provides a control with the same channel assignments as interleave but a
different subarray distribution.

For 64 pages, one entry per page and stride~32, active-page striping reduces
kernel time from 1,092,373 to 277,353 cycles (3.94$\times$ speedup), with
64 array reads in both runs. READ traces show that striping spreads accesses
over four subarrays instead of one, allowing more reads to proceed in parallel.
At stride~4, the same policy change increases time from 141,530 to 277,353
cycles. Although requests become balanced across channels, they use four
subarrays instead of eight, reducing the available parallelism.
Table~\ref{tab:placement_tradeoff} shows the same trend across both working-set
sizes and entry counts. With stride~1, both mappings use at least eight
subarrays, enough to reach the eight-operation cap, and their execution
times are similar.

\begin{table}[t]
\centering
\setlength{\abovecaptionskip}{10pt}
\setlength{\belowcaptionskip}{-10pt}
\footnotesize
\setlength{\tabcolsep}{5pt}
\begin{tabular}{@{}rrrrr@{}}
\toprule
Pages & Stride & Subarrays & \multicolumn{2}{c}{Kernel speedup} \\
 & & I / S & $E=1$ & $E=8$ \\
\midrule
64 & 1 & 32 / 8 & 1.00 & 1.00 \\
64 & 4 & 8 / 4 & 0.51 & 0.51 \\
64 & 32 & 1 / 4 & 3.94 & 3.93 \\
128 & 1 & 32 / 8 & 1.06 & 1.00 \\
128 & 4 & 8 / 4 & 0.51 & 0.51 \\
128 & 32 & 1 / 4 & 3.97 & 3.97 \\
\bottomrule
\end{tabular}
\caption{\textbf{Active-page striping (S) versus built-in interleave (I) under
default timing.} Speedup is kernel cycles of I divided by S. Values below one
indicate a slowdown. $E$ is the number of 128\,B entries per page. Subarray
counts are distinct resources used over a run, with identical counts for both
$E$ values. The concurrent-operation cap is eight.}\vspace{-2pt}
\label{tab:placement_tradeoff}
\end{table}

\begin{figure*}[t]
\centering
\setlength{\abovecaptionskip}{-5pt}
 \setlength{\belowcaptionskip}{-10pt}
\includegraphics[width=0.9\textwidth]{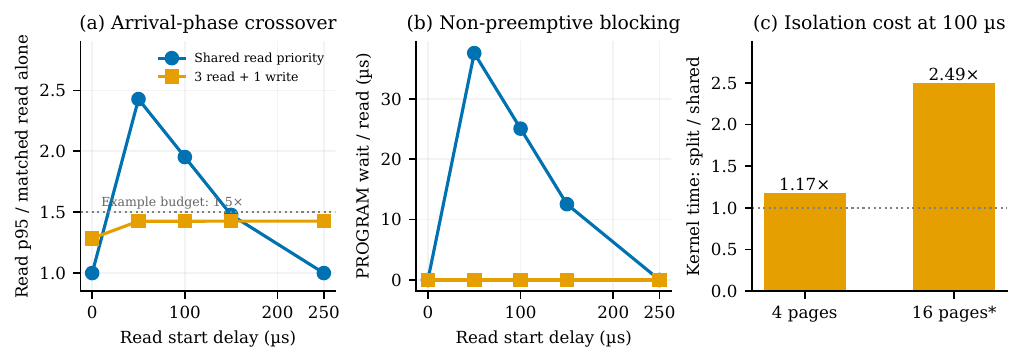}
\caption{\textbf{Read protection depends on arrival phase and resource allocation.}
(a) Read p95 is normalized to a matched read-only run.
(b) Average time per read spent queued while its target subarray is executing
PROGRAM. The average includes all read requests and is not added to kernel time.
(c) Isolation changes whole-kernel time. The starred 16-page comparison uses
a matched, extended 400\,$\mu$s assembly deadline. The original isolated
configuration is infeasible under the shorter deadline. The 1.5$\times$ line
is an illustrative budget.}
\label{fig:case_isolation}
\end{figure*}

\ddparagraph{Design implication.}
Page placement can lower read-side page-service amplification.
Its effect on execution time also depends on array work and the parallelism
available across subarrays. Placement policies should therefore consider
page-service reuse together with channel and subarray assignments.
HBF-Sim's mapping interface and request traces make both effects visible
when evaluating static placement policies.

\subsection{Protecting Reads from Write Interference}
\label{sec:write_isolation}

\ddparagraph{Question and policies-2:}
\textit{When reads arrive during a long flash program, does read-priority scheduling
provide adequate protection, or should reads receive separate resources?}
We compare shared FCFS, shared read priority, shared write drain, and static
1R/3W, 2R/2W and 3R/1W channel partitions using read priority within the
partitioned configuration. Scheduling changes candidate selection. Partitioning
changes the page-to-channel map. Neither interrupts an issued PROGRAM.
Read and write pages are disjoint, and all configurations retain strict
full-page writes and acknowledgements after programming.

\ddparagraph{Matched experiment.}
The phase study holds four channels, four total subarrays and four active slots
fixed. It varies writes of 4/16 pages and the nominal read-start delay, with
matched read-only controls. Of 37 configurations, 36 complete and one violates
the strict page-assembly deadline. The main deadline is 10,000 DRAM ticks
(11.76\,$\mu$s). The matched extended-deadline control uses 400\,$\mu$s.
The read stream contains 512 sector requests, and each written page contributes
128 sectors with complete byte coverage. We report creation-to-controller-return
latencies for reads and writes, together with whole-kernel completion time.

\ddparagraph{Read benefit and write cost.}
With four written pages and a 100\,$\mu$s read-start delay, 3R/1W reduces
read p95 from 232,456 to 169,730 GPU cycles relative to shared read priority
(27.0\%). Kernel time increases from 776,097 to 911,850 cycles (17.5\%),
while write p95 increases from 226,715 to 905,889 cycles (4.00$\times$).
The reserved read resources reduce exposure to programming, while the single
write channel limits write parallelism.

The benefit changes with arrival phase (Figure~\ref{fig:case_isolation}).
At zero nominal delay, 3R/1W increases read p95 from 118,999 to 152,968 cycles.
At 50\,$\mu$s it lowers p95 from 289,058 to 169,710 cycles. At 250\,$\mu$s,
shared read priority again has lower read p95. Priority can reorder pending
work, but an already-issued program leaves residual blocking. Nominal launch
delay and actual program overlap are therefore evaluated separately using
request and media events.

\ddparagraph{Design implication.}
HBF channel allocation should use read/write isolation selectively, balancing read-tail latency targets against write parallelism and whole-workload completion time.
For an illustrative budget of read p95 at most 1.5$\times$ the matched
read-only p95, the four-page sweep gives a concrete choice between shared
read priority and 3R/1W. At 0, 150 and 250\,$\mu$s, sharing satisfies the
budget and finishes the kernel sooner. At 50 and 100\,$\mu$s, only 3R/1W
of these two options meets the budget. 
The 16-page deadline failure further shows why strict-write feasibility must
be checked before ranking performance. 
%

\section{Discussion}
\label{sec:disc}



HBF-Sim models page mapping, read coalescing, write aggregation, channel contention, and completion feedback to GPU execution. 
ECC, read retry, refresh, and thermal effects are not modeled in the current version. 
These mechanisms may affect tail latency and sustained bandwidth in real devices, but they do not invalidate the functionally consistent request/completion behavior or the semantically correct page-level interactions captured by our simulator. 
Since HBF hardware is unavailable, we check selected properties from the OCP specification~\cite{ocp_hbf_v07} without hardware-calibrated measurements: requests access only channel-local resources, programming requires a complete 4\,KiB page, and write acknowledgements follow programming (Section~\ref{sec:validation}). 
We also perform timing experiments to examine page-service and resource-scaling trends. 
These evaluations demonstrate that HBF-Sim faithfully preserves the OCP-specified request/completion semantics, though they do not establish cycle-level accuracy against physical HBF silicon.

Future work can extend the static policies evaluated here to adaptive
placement and channel allocation based on access patterns. Transaction-level
AXI/UCIe modeling would allow closer examination of protocol behavior beyond
the current bandwidth-and-credit model. Hardware measurements, when
available, could calibrate timing parameters and test whether the observed
trade-offs hold on real devices.

\vspace{-5pt}
\section{Related Work}
\ddparagraph{Device and interconnect models.} Xerxes~\cite{xerxes} studies scalable CXL fabrics, CXL-MQSim~\cite{cxl_mqsim}
studies CXL-enabled SSDs, and WARP~\cite{warp} combines device emulation with
real-device characterization. SwarmIO~\cite{swarmio} emulates SSDs for GPU-initiated I/O.
Sano et al.~\cite{gpu_cxl_graph} use FPGA prototypes to study GPU graph processing with microsecond-latency CXL memory. FlashSim~\cite{flashsim} and
NANDFlashSim~\cite{nandflashsim} model flash timing and media operations.
Ramulator~\cite{ramulator} and DRAMsim3~\cite{dramsim3} model DRAM devices.
These models address complementary device or interconnect questions. Their
interfaces and media-management assumptions require examination when reused
for an HBF tier.

\ddparagraph{HBF architecture and systems studies.} H${}^3$~\cite{h3} explores hybrid HBM--HBF inference, and
HAVEN~\cite{haven} targets approximate nearest-neighbor search.
TileLens~\cite{tilelens} studies memory layout and prefetching for
large-granularity memory. FlashAccel~\cite{flashaccel} co-designs HBF hardware
and software for LLM inference. FLINT~\cite{flint} studies HBF-based inference
and read coalescing. The full-stack characterization~\cite{hbfsucks} and
IEEE CAL study~\cite{hbf_llm_cal} examine HBF deployment trade-offs.
Petrucci et al.~\cite{hbf_all_you_need} compare HBF and LPDDR capacity tiers using a roofline model.
These architecture and application studies motivate reusable evaluation
infrastructure.
While these studies demonstrate HBF's potential in specific workloads, their models lack closed-loop timing feedback between GPU execution pipelines and HBF media constraints.


\section{Conclusion}

This paper addresses the lack of an open and faithful simulation
platform for exploring high-bandwidth flash in next-generation GPU
big-memory systems. We present \textbf{HBF-Sim}, an extensible
Accel-Sim-integrated framework that models the full
GPU--HBF request/completion path for systematic design exploration.
Our results show that HBF-Sim is functionally correct end to end,
captures closed-loop effects that simpler models miss, and
reproduces expected page-service and resource-scaling trends.
Using HBF-Sim, we further evaluate the impact of page placement
and resource isolation on performance, and identify two design
implications for HBF systems. HBF-Sim thus provides an open
foundation for reproducible HBF research before production
hardware becomes available.

\bibliographystyle{plain}
\bibliography{hbf-sim}

\end{document}